# VEHICLE ELECTRIFICATION SOLUTIONS: review and open challenges

Youssef El Amrani[1]*, Saad Motahhir[2], Abdelaziz El Ghzizal[1]

[1]Innovative Technologies Laboratory, EST, SMBA University, Fez, Morocco

[2]ENSA, SMBA University, Fez, Morocco

*youssef.elamrani7@usmba.ac.ma

## Abstract

An Electric Vehicle usually refers to any vehicle that is partially or fully powered by a battery that can be directly plugged into the mains. Therefore, the new vehicles provide various benefits, including convenience, efficiency, sustainability, and economy. The present study concerns a comprehensive review of vehicle electrification solutions. Indeed, the major electric vehicle technologies are presented. Moreover, based on several research works from the literature, the main electrification solutions are illustrated, including degree of electrification, battery system management, onboard chargers, and power converters. In addition, these solutions, as well as the open challenges, are discussed and evaluated.

# 1. Introduction

The world is facing pressure on fossil fuels, due to high demand and a large amount of $CO_2$ emissions and especially for the high demand for vehicles, According to the International Energy Agency; Figure 1 depicts the proportional contribution of $CO_2$ emissions in transportation as well as other sectors, that transportation is one of the major emissions sectors accounting for 22 percent of total $CO_2$ emissions in 2020 **[1,8]**.

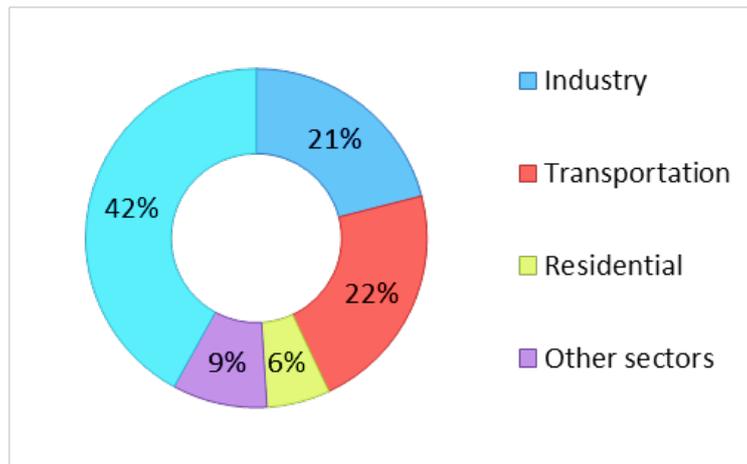

**Figure 1.** $CO_2$ emissions by various sectors.

Thus, most countries are seeking to move towards economic, sustainable, reliable, efficient, and green energy resources, so vehicle electrification can be a significant solution in this regard to using electricity to power the vehicle, transferring vehicle components that use traditional energy source with components that use electricity **[1]**.

Electric vehicles are gaining popularity as a result of global trends such as advances in technology and the commercial viability of renewable energy technologies, as well as the development of electric motors and electronic control systems, particularly supporting technologies such as Grid-to-Vehicle (G2V) and Vehicle-to-Grid (V2G), which offer numerous advantages over traditional vehicle technologies presented in zero-emission, simplicity, reliability, efficiency, accessibility, so on **[1,20]**.

To resolve those problems, vehicle electrification can be an excellent alternative and has been the subject of several studies presented in the battery management system, power converters, onboard chargers so on **[14]**.

During this context, electrification will take multiple forms, and does not essentially mean entirely battery electrical vehicles, however, the bulk of the vehicle power comes originally from electricity, and thus might even be plug-in-hybrids or cell vehicles **[14,20]**.

Contrary to previous research reviews, this work summarizes and combines different subjects, including the technology, components, and solutions for electrification, as well as the same challenges presented in electric vehicles that can facilitate the understanding of the topic.

This review contributes to a good and current understanding of the types, technology, and characteristics of electric vehicles. Then show clearly the levels of electrification, and main electrification solutions and give the researchers a reference point to where they can start their improvements.

The rest of this paper is structured as follows: In part 2, we shall carry out the research approach that we used to create this review, part 3 categorize electric vehicle technologies by discussing the types, characteristics, and levels of electrification, and part 4 reviews the main electrification solutions. Then the open challenges are covered in part 5. Finally, part 6 concludes this review.

## 2. Research methodology

To achieve the objectives of our work and to analyze the existing studies in the literature, we forced on the data sources mentioned below.

- Springer
- ScienceDirect
- Google Scholar
- IEEE Xplore
- Wiley Online Library

## 3. Electric Vehicle Technologies

In this part, we give a taxonomy of the main types of EVs, along with comments on their primary qualities. We also talk about the degree of EV electrification based on the various technologies used for EVs.

### 3.1. Electric VEHICLES Taxonomy

Depending on the technology used in their engines, we can now find many sorts of electric vehicles. They are classified into five groups as given in Figure 2 **[2,3,4,5,6,7]**.

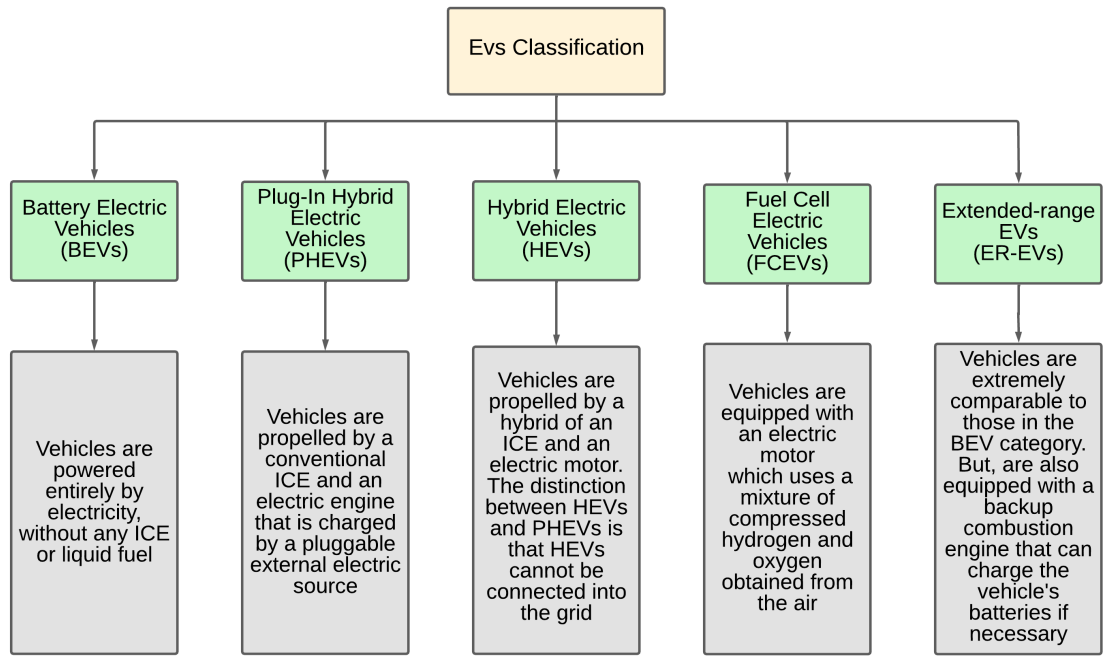

**Figure 2.** Classification in EVs.

## 3.2. Vehicle electrification

### A. Definition

The concept of electrification includes using so much electrical energy to power vehicle propulsion and non-propulsion loads. ICEs are generally inefficient, with an average efficiency of less than 30%. Electric systems, on the other hand, can supply substantially higher efficiency. Electric motors can be made to operate at efficiencies of more than 90% **[8]**.

Moreover, electrical systems are quicker and simpler to control than mechanical systems. Additionally, electrical energy can be created from a variety of renewable and carbon-free resources, such as wind, sun, and hydro **[9].**

### B. Degree of Electrification

EVs can be classified as having varying degrees of electrification, ranging from nothing, just an internal combustion engine (ICE), to completely electrified with only a motor and no engine **[10].**

Vehicle technology is often categorized into four main areas **[11,12,20]**, as indicated in Figure 3, with electrification increasing from left to right. The energy sources of a traditional vehicle (ICE) are gasoline or diesel fuel, both of which are significant emitters of carbon dioxide to the environment. As a result, hybrid automobiles emit less carbon than gasoline-powered vehicles. The fuel cell vehicle and electric vehicle categories are also well known as zero-emission vehicles since they rely on hydrogen fuel cells and batteries (measured at the tailpipe).

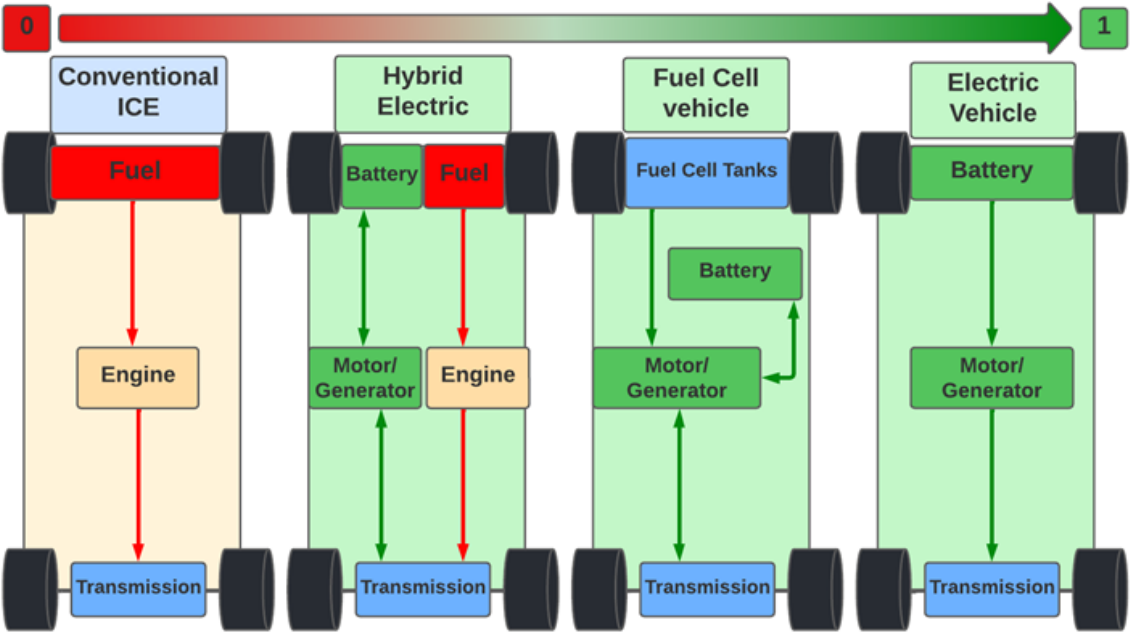

**Figure 3.** Degree of Electrification of EVs.

## 4. Vehicle electrification solutions

Electric vehicles are generally used for pollution-free transportation. However, it has been viewed that the distance traveled by a battery-powered electric vehicle is significantly less than that of a fuel-powered engine, and there is insufficient regenerative energy captured from the vehicle's kinetic energy. There are numerous sorts of losses in power converters that affect battery energy consumption. To increase the distance traveled by electric vehicles and the amount of regenerative energy captured, we must improve the performance of all components used in electric vehicles, such as the electric motor, energy storage system (battery, supercapacitor...), and power converter (DC/DC, AC/DC...), as shown in figure 4. This section reviews the major components used in an electric vehicle. **[13].**

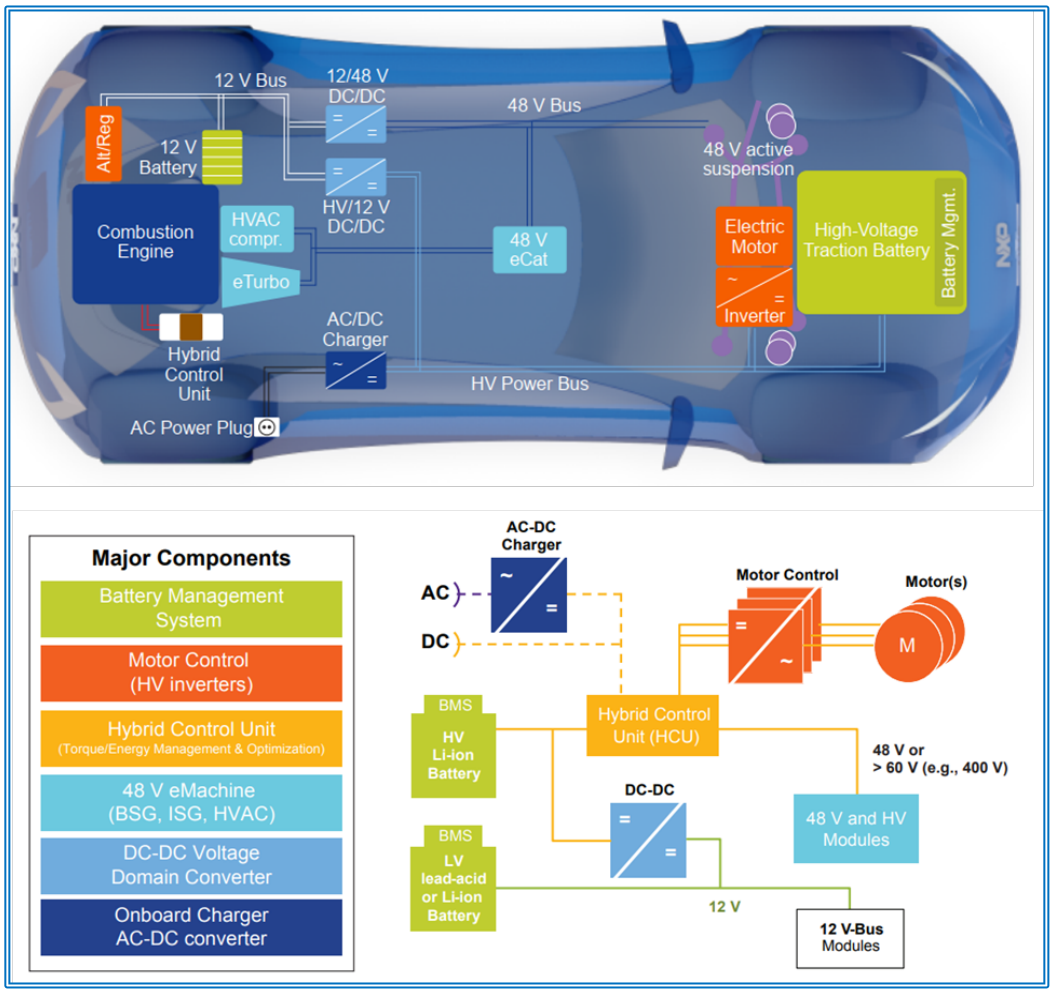

**Figure 4.** Block Diagram and key component of an electric vehicle. **[14].**

## 4.1. Battery Management System

The Battery Management System (BMS) is the heart of an electric vehicle. The battery management system is the basic device connected between the charger and the battery of electric or hybrid systems. The battery management system presents many functionalities and technologies and is essential for guaranteeing safety, protection, monitoring, communication, charging and discharging management, and cell balancing. Figure 5 classified the major BMS functionalities, and Figure 6 showed the relation of key technologies in the BMS **[15].**

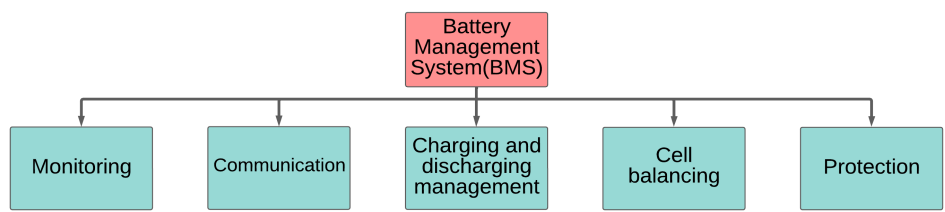

**Figure 5.** BMS functionalities.

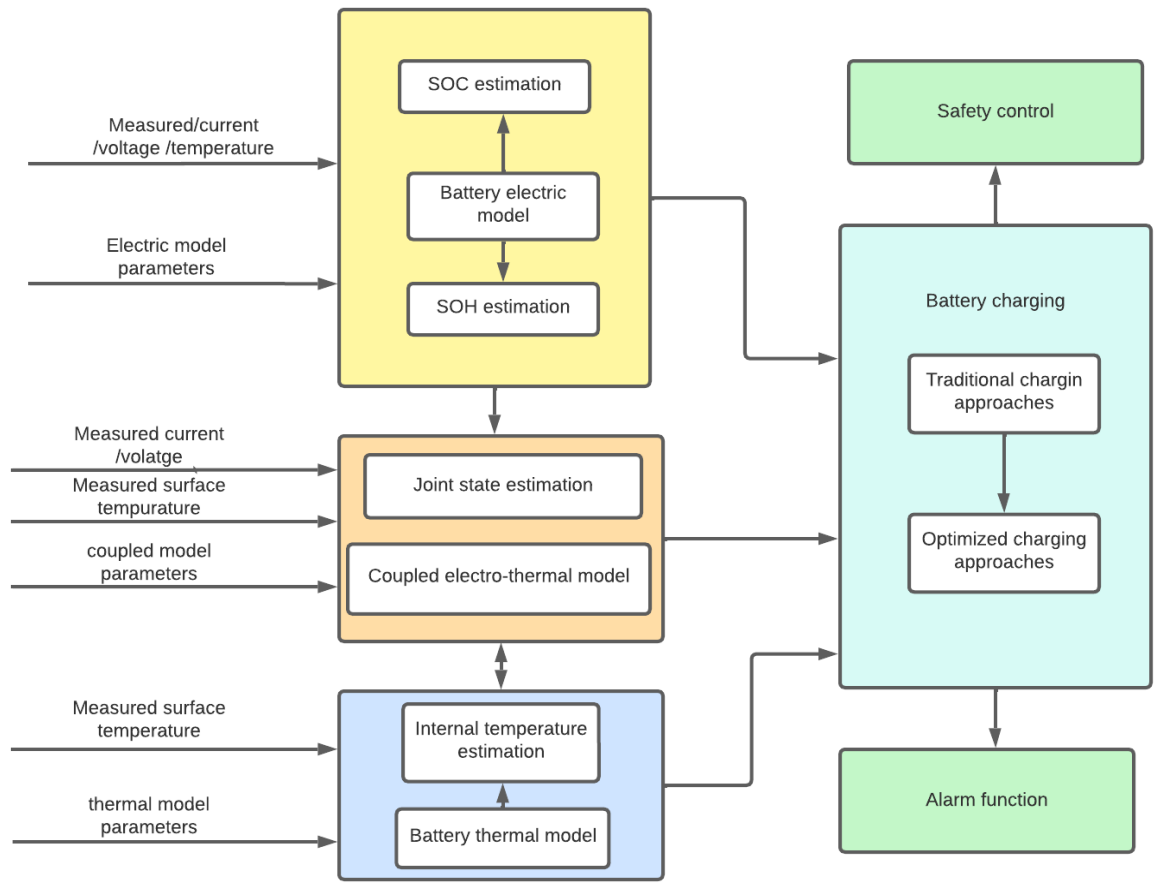

**Figure 6.** The relation of key technologies in the BMS.

## 4.2. BMS for Electric Vehicle (EV)

The BMS is critical to the overall command of an EV. Figure 7 displays the powertrain system structure of a battery-powered EV. The traction battery, which has a large capacity and high power, is the only source of power. It operates in two different styles, charging and discharging. While in discharge mode, it powers the electric motor, which converts electrical energy to mechanical energy. The mechanical drive delivers rotational energy to the wheels of the vehicle. **[16].**

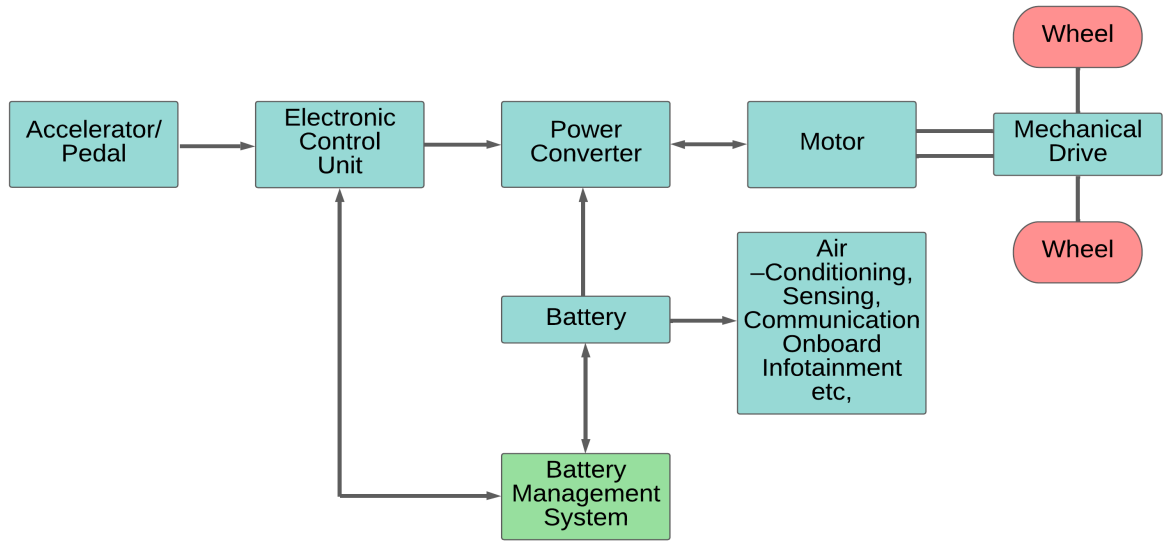

**Figure 7.** BMS operation inside Electric Vehicle (EV).

## 4.3. Electric vehicle DC-DC converters

To connect the battery, FC, or other technology to the DC-link, a DC-DC converter is necessary. The DC-DC is a kind of power converter that converts a direct current (DC) source from a single level of voltage to another by temporarily conserving the input of energy and then releasing it to a different voltage level at the output. Figure 8 depicts the power of an electric drive system, whereas Figure 9 depicts the most popular DC-DC converters **[17]**.

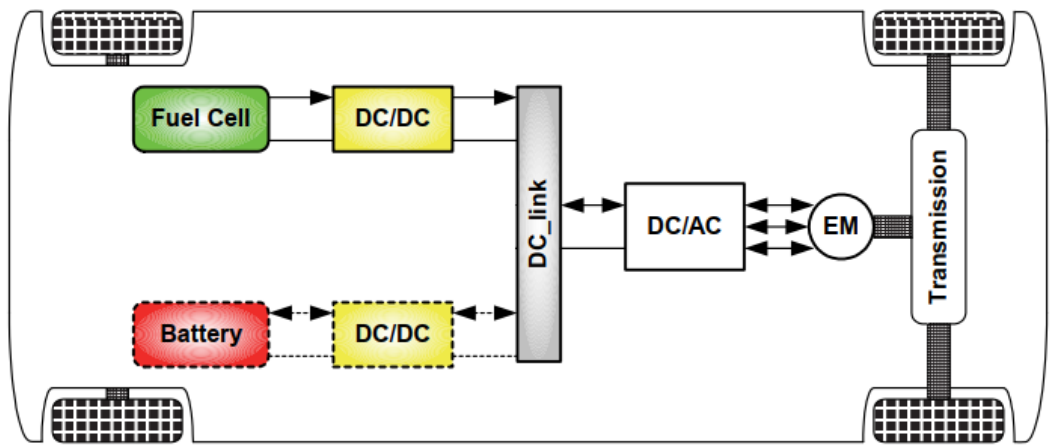

**Figure 8:** DC/DC converters in EV.

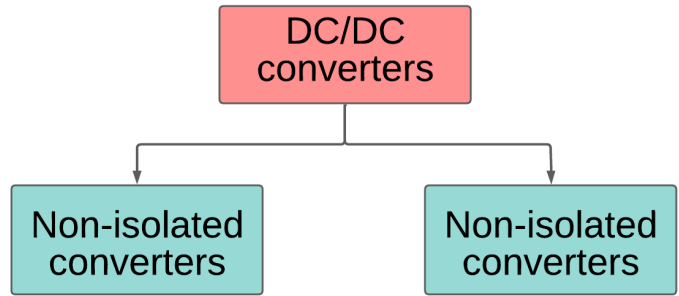

**Figure 9:** Most common DC/DC converters.

There are five types of converters in this non-isolated category which are: boost, buck, cuk, buck-boost, and charge-pump converters. The boost converter increases the voltage. Whereas the buck converter decreases in voltage, the cuk and buck-boost can step up and step down. The charge pump is designed to invert or step-up voltage.

In this isolated category, a high-frequency transformer is used in these types of converters. It is necessary to use an isolated converter in applications where the output needs to be completely isolated from the input. Within this category, there are a lot of types of converters, including Push-Pull DC/DC, Fly-back, Full-Bridge, and Half-Bridge converters. These converters can all be used as bidirectional converters, and the voltage stepping ratio is high **[17].**

## 4.4. On-Board Chargers

The onboard charger (OBC) allows EVs to be charged directly from the AC grid and is commonly utilized in the automobile industry due to its convenience, particularly when compared to the high cost and high volume of off-board charging alternatives. Figure 10 depicts a comprehensive representation of the OBC with the various converters, as well as the battery and the grid **[18].**

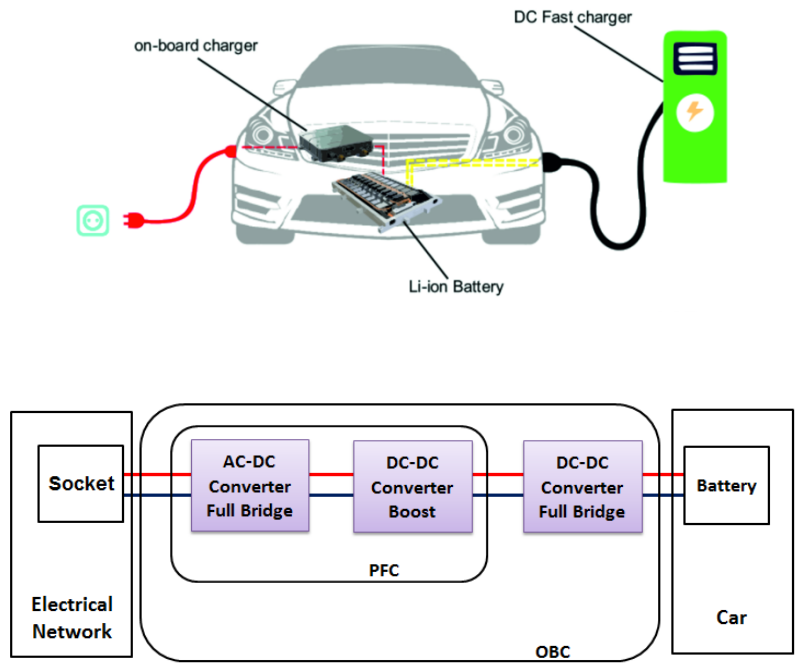

**Figure 10.** General diagram of the onboard charger.

A power factor corrector (PFC) and a DC-DC converter comprise the onboard charger. A PFC is an AC-DC converter that combines an AC-DC with a DC-DC boost converter. A PFC's objective is to extract a sinusoidal waveform current from a network. PFC control requires a voltage sensor to detect the load voltage and a current sensor to measure the inductor current in the boost converter. With this understanding, you can manipulate the PFC. **[18].**

## 4.5. Hybrid Control Unit

The hybrid control unit (HCU) is an important part of hybrid and electric vehicles control systems. It calculates and manages output characteristics, for instance, engine power and motor torque, using input signals **[14].**

## 5. Open Challenges

In this section, we discuss aspects that are still open or interesting to be investigated to propose new and improved solutions in EVs.

### 5.1. Batteries Technologies

Batteries are one of the most important components of EVs because they continue to be one of the most expensive components of the overall car cost. As a result, batteries have a direct impact on EV performance. Advances in durability, charging density, and charge and discharge processes necessitate the employment of different resources in the development of new technologies capable of outperforming lithium-ion batteries, which are widely utilized in the automotive industry **[19,21].**

### 5.2. Communications in electric vehicles

Vehicular communications can help to accelerate sustainable modes of transportation. Wireless communication networks lead automobiles to be outfitted with a communication system that will enable vehicle-to-vehicle (V2V) and infrastructure connections (V2I) **[19,21].**

### 5.3. Artificial intelligence in electric vehicles

Artificial intelligence (AI) is also will assist to accelerate the automation of electric vehicle driving. The usage of AI-based algorithms will bring intelligence to cars, opening up a lot of new potentials that will change future transportation systems. We can find multiple AI-based solutions connected to several EV domains, such as energy-efficient routing, smarter charging, and charging electric vehicles with renewable energy-generated electricity or battery temperature management **[19,21].**

## 6. Conclusion

The use of electric vehicles for transportation is regarded as an essential component for managing sustainable development and environmental challenges. In this review article, we began with a general introduction to the context of our subject, followed by the EV taxonomy, the degree of electrification, the major components used in an EV, and finally, the open challenges that EVs must face. Regarding the EV taxonomy, there are five main taxonomies: BEVs, PHEVs, HEVs, FCEVs, and ER-EVs. We analyzed the differences between them by discussing the characteristics, benefits, and uses of each. For the degree of electrification, we have classified these types of EVs by order. Concerning the major components, we have shown the link between all the main components used in EVs and the characteristics of each of them. The BSM is dedicated mainly to safety and cell balancing. Regarding the onboard charger, it acts as a link between the electrical grid and the battery in an EV. The hybrid control unit aims to calculate and manage the output parameters using the input signals.

Finally, the challenges that EVs must overcome, essentially in batteries and new technologies based on AI, are discussed. Because it covers combination technologies, this review could be a useful reference to help researchers in learning about the electrification topic.

# Nomenclature

| | |
|---|---|
| **EV** | Electric vehicles |
| **$CO_2$** | Carbon Dioxide |
| **G2V** | Grid-to-Vehicle |
| **V2G** | Vehicle-to-Grid |
| **BEV** | Battery Electric Vehicles |
| **HEV** | Hybrid Electric Vehicles |
| **HyEV** | Hydrogen Electric Vehicles |
| **PHEV** | Plug-In Hybrid Electric Vehicles |
| **FCEV** | Fuel Cell Electric Vehicle |
| **ER-EV** | Extended-Range Electric Vehicles |
| **ICE** | Internal Combustion Engine |
| **BSM** | Battery System Management |
| **SOC** | State of Charge |
| **SOH** | State of health |
| **OBC** | On Board Charging |
| **PFC** | Power Factor Corrector |
| **HCU** | Hybrid Control Unit |
| **AI** | Artificial Intelligence |
| **V2V** | Vehicle-to-Vehicle |
| **V2I** | Vehicle-to-Infrastructure |
| **AC/DC** | Alternating Current/Direct Current |